\documentclass[hyper]{prop2015}
\usepackage[english]{babel}
\usepackage{amsthm}
\usepackage{bbm}
\newcommand{\dd}{\mathbf{d}}
\theoremstyle{plain}
\newtheorem{theo}{Theorem}[section]

\theoremstyle{definition}
\newtheorem{defi}[theo]{Definition}

\category{Proceedings}
\subtitle{\href{http://www.maths.dur.ac.uk/lms/109/index.html}{LMS/EPSRC Durham Symposium on Higher Structures in M-Theory}}
\title{Homotopy Algebras in String Field Theory}
\author[IS]{Ivo Sachs\inst{a,}\footnote{Corresponding author e-mail:~\href{mailto:Ivo.Sachs@lmu.de}{\textsf{Ivo.Sachs@lmu.de}}}}
\address[1]{Arnold Sommerfeld Center for Theoretical Physics, Ludwig Maximilian University of Munich,
Theresienstr. 37, D-80333 M\"unchen, Germany}
\begin{abstract}
Homotopy algebra and its involutive generalisation plays an important role in the construction of string field theory. 
I will review recent progress in these applications of homotopy algebra and its relation to moduli spaces. 
\end{abstract}
\shortabstract
\begin{document}
\maketitle

\section{Introduction}
String field theory (SFT) \cite{Witten:1985cc,Zwiebach:1992ie} comes naturally with the structure of  Batalin--Vilkoviski (BV) action (e.g.~\cite{Henneaux:1987cp}). In particular, anti fields arise simply by relaxing the world-sheet ghost number for the states in the world-sheet CFT. On the other hand, the standard construction of SFT is rather different from the usual treatment of BV actions which starts from a given action $S_0$ at ghost number zero. Its  extension to fields of non-zero degree is then determined by the local symmetries of $S_0$. Alternatively, one can start with a geometric setting which consists of a space of fields, $\mathcal{F}$ and a homological vector field $Q$ (e.g.~\cite{Cattaneo:2012qu}). In contrast, in string theory, $S_0$ is not known and neither is $\mathcal{F}$ nor $Q$. However, a BV structure arises geometrically from the decomposition of the moduli space of the world-sheet (punctured Riemann surfaces) of perturbative string theory. Below we will describe how these two concepts come together when a BV action is expanded around a chosen point $p$ in  $\mathcal{F}$ which is the usual setting of SFT. When expanded in this way both constructions give rise to homotopy algebras (or $\infty$ structures).  

In Section~\ref{sec:2}, we first review the geometric approach to BV quantization, in particular its expansion in $T_p\mathcal{F}$. In Section~\ref{sec:3}, we will see how the same algebraic structure arises in SFT. Finally, in Section~\ref{sec:4}, we will review how this algebraic structure can be used to construct superstring field theory where neither the BV action nor the geometric BV structure is known. 

\section{BV actions}\label{sec:2}
A consistent choice of BV data is given by the triple $(\mathcal{F}, \omega, V)$, where $\mathcal{F}$ is a graded space of fields, $\omega$ is a non-degenerate, odd symplectic structure and $Q$ is a homological vector field which defines a cohomology on the functionals on $\mathcal{F}$. This cohomology represents gauge invariant functionals modulo gauge equivalences. One can transfer this cohomological structure to the tangent space $T_p\mathcal{F}$ by a Taylor expansion
\begin{equation}\label{VQexp}
Q_p = Q_p^{(0)} + Q_p^{(1)} + \cdots\, .
\end{equation}
around a point $p$ in field space. The $Q_p^{(n)}$ define multilinear maps
\begin{equation}
Q_p^{(n)}: (T_p \mathcal{F})^{\otimes n} \rightarrow T_p \mathcal{F}.
\end{equation}
If $p$ is a critical point in $\mathcal{F}$, then $Q_p^{(0)} = 0$.  The maps $Q_p^{(n)}$ together with $\omega_p$ then define the vertices of order $n+1$ of a field theory action $S$ expanded around the point $p$. 
These vertices define an infinity structure ($L_\infty$- or $A_\infty$-algebra) on $T_p \mathcal{F}$. 

The differential
\begin{equation}
\mathbf{d}\equiv Q_p^{(1)} : T_p \mathcal{F} \rightarrow T_p \mathcal{F}
\end{equation}
in (\ref{VQexp}) encodes the free equations of motion as well as the linear part of the gauge transformations of the theory expanded around the point $p$. The cohomology $H = H(T_p \mathcal{F},Q_p^{(1)})$ is given by the equivalence classes of on-shell fields modulo linear gauge transformations. Typically, they correspond to asymptotic states of the theory.

One can transfer the $\infty$ structure on $T_p\mathcal{F}$ to a {\it minimal} $\infty$ structure on $H$, essentially by evaluating the Feynman diagrams with the vertices  $Q_p^{(n)}$. We denote the multilinear maps of the minimal algebra by ${Q}_{\text{min}}^{(n)}: H^{\otimes n} \rightarrow H$, where now, in addition,  $Q^{(1)}_\text{min} \equiv 0$. At ghost number zero they represent the tree-level S-matrices, in the sense that the $n$-point scattering of states $x_1,\ldots,x_n$ is equal to
\begin{equation}
\mathbf{S}^{(n)}(x_1,\ldots,x_n) = \frac{1}{n}\omega_p\left(x_1,Q_\text{min}^{(n-1)}(x_2,\ldots,x_n)\right).
\end{equation}

We can summarize the above steps in the following diagram
\begin{eqnarray}
(\mathcal{F}, \omega, Q)& \stackrel{\text{crit. point}}{\longrightarrow}& (T_p\mathcal{F},\omega_p,\{Q_p^{(n)}\}) \\&&\stackrel{\text{min. model}}{\longrightarrow} (H(T_p\mathcal{F},Q_p^{(1)}), \omega_p,\{Q_{\text{min}}^{(n)}\}).\nonumber
\end{eqnarray}

In string theory this construction was developed backwards, and also only partially so. The original string theory is a theory of S-matrices only. It is defined on-shell (i.e. on $H(T_p\mathcal{F},Q_p^{(1)})$). String field theory is then, essentially, the reconstruction of $Q_p^{(n)}$ given ${Q}_{\text{min}}^{(n)}$ and $\dd$. As such it is defined by a triple $(T_p\mathcal{F},\omega_p,\{Q_p^{(n)}\})$. In particular, this construction relies on a, a priori choice, of a background, that is, a critical point $p\in \mathcal{F}$.

Before we close this section we would like to point out that with the data specified in this section one can always infer the existence of a BV action. Indeed given $Q$ and $\omega$ as above on can define an exact differential 
\begin{equation}
\dd S:=i_Q\omega
\end{equation}
so that $S$ satisfies the classical BV master equation $(S,S)=0$.

\section{String field theory}\label{sec:3}

In bosonic string field theory the starting point is a different one. This is because we do not know how to describe  $\mathcal{F}$ let alone $Q$, except at some very special points in $\mathcal{F}$. On the other hand, the world-sheet approach computes on-shell amplitudes as conformal field theory correlation functions on punctured Riemann surfaces, $\Sigma_{g,n}$ together with a coordinate curve around each puncture.  The corresponding moduli space $\hat P_{g,n}$ is a bundle over $P_{g,n}$, is the moduli space of punctured Riemann surfaces of genus $0$ with $n$ punctures of complex dimension $3g+n-3$. For the sake of simplicity we will focus on genus zero, that is classical-, or tree level SFT in what follows. A consistent decomposition of its moduli space $\hat P_{n}$ satisfies the geometric BV (or master) equation \cite{Zwiebach:1992ie,Zwiebach:1997fe}
\begin{align}\label{BVgcq}
&\partial \nu_{n}+\frac{1}{2}\sum\limits_{n_1\leq n_2}(\nu_{n_1+1},\nu_{n_2+1})=0&\,,
\end{align}
where $n=n_1+n_2$.  The geometric vertices, $\nu_{n}$ with labeled punctures are elements in the singular chain complex $C^{\bullet}(\hat{P}_{n})$ endowed with an orientation. The grading is defined by the co-dimension, 
\begin{align}
\text{deg}(\nu_{n}) = \text{dim}({P}_{n})-\text{dim}(\nu_{n}) \,.
\end{align}
With this grading the boundary operator $\partial$ has degree one. 
The BV bracket of two geometric vertices is defined geometrically in terms of the glueing of coordinate discs around one puncture of each vertex. 

The world-sheet CFT provides a BV morphism from this chain complex to the endomorphism operad  End$_V$, where $V$ is the state space of the (matter and ghost) CFT. The family of such morphisms is conveniently paramet\-rized by the CFT correlation functions (e.g. \cite{Witten:2012bh})
\begin{align}
\alpha_{x_1,\ldots,x_{m+3}}(h,\delta h)=\langle e^{-i<\delta h, b>}\prod\limits_{j=1}^{m+3}( f^*_j(\mathbf{V}_j))(0)\rangle\,,
\end{align}
where $x_j\in V$, $ m=\text{dim}(\nu_{n})$,  $h$ determines the conformal structure on the world-sheet and $\delta h$ is taken to be the variation of the conformal structure (with opposite parity assignment) and which is paired with $b$, the symmetric and traceless odd world-sheet tensor that originates in the standard BRST quantization of the world-sheet Polyakov action. Finally, $f^*_j(\mathbf{V}_j)$ is the pull back of the world-sheet conformal field (or vertex operator) at the puncture $z_j$ to the origin of the complex plane by the conformal mapping $f_j$ (which depends in $h$). As such, $\alpha_{x_1,\ldots,x_{m+3}}$ defines a function on the twisted tangent bundle, $\Pi T\hat P_{n}$. 

In order to integrate $\alpha_{x_1,\ldots,x_{m+3}}$ over $\Pi T\hat P_{n}$ one can choose a suitable parametrisation $\{t_j\}$, $j=1,\ldots,m$, of the subspace $\nu_{n}$ of ${\hat P}_{n}$ such that
\begin{align}
\delta h=\sum\limits_{j=1}^{\text{dim}(\nu_{g,n})}\frac{\partial h}{\partial t_j}\dd t_j\,.
\end{align}
Then, integrating over the odd variable $\dd t_j$ produces a differential form $\tilde\alpha_{x_1,\ldots,x_{m+3}}$  of degree $m$ that can be integrated over a section of $\hat P_n$, i.e. a top form on ${P}_{n}$. Integration over ${\nu}_{n}$ then defines a multilinear function on $V$. We can go further using that the BPZ inner product in CFT gives rise to a non-degenerate, odd symplectic form $\omega$ on $V$. Thus we can write
\begin{equation}
\begin{aligned}
\int\limits_{\nu_n}\tilde\alpha_{x_1,\ldots,x_{m+3}}&=: C_{m+3}(x_1,\ldots,x_{m+3})\\&:=\omega(x_1,l_{m+2} (x_2,\ldots,x_{m+3}))\,,
\end{aligned}
\end{equation}
where $l_n\in {\rm End}_V $.
 
To complete the description of this morphism we still need to define the image of the boundary operator $\partial$. For this we note that the reparametrization invariance of the world-sheet theory implies the existence of a Noether current 
$j_{\small{BRST}}\in \Omega^1(\Sigma_n,{\rm End}_V)$ with corresponding charge 
\begin{equation}
\oint_C j_{BRST}\;=\;Q_0 \;\in \;\Omega^1({\rm End}_V)\,.
\end{equation}
Furthermore, it can be shown (e.g. \cite{Witten:2012bh}), that $\alpha\circ Q_0=\dd\alpha$ so that 
\begin{eqnarray}
\int\limits_{\nu_n}\tilde\alpha_{(Q_0 x)}&=&\int\limits_{\nu_n}\dd \tilde\alpha_{ x}= \int\limits_{\partial \nu_n}\tilde\alpha_{ x}\\&=&-\frac{1}{2}\sum\limits_{n_1\leq n_2}\int\limits_{(\nu_{n_1+1},\nu_{n_2+1})}\tilde\alpha_{x}\nonumber\\
&=&-\frac{1}{2}\sum\limits_{n_1\leq n_2}\left(\int\limits_{\nu_{n_1+1}}\tilde\alpha, \int\limits_{\nu_{n_2+1}}\tilde\alpha\right)(x)\nonumber
\end{eqnarray}
where, in the last equality, the bracket stands for the algebraic BV  bracket rather than the geometric bracket in the second line. This then proves that the world-sheet CFT realizes a morphism between BV algebras. In sum, 

\smallskip
\begin{defi}
Classical bosonic string field theory is a BV morphism from the set of Maurer--Cartan elements of the chain complex of geometric vertices (subspaces) of the moduli spaces, $\hat P_n$ of punctured spheres with parametrized curves, to the set of Maurer--Cartan elements of the algebraic BV equation on End$_V$.
\end{defi}

\smallskip

\noindent An immediate consequence is then that $l_1:=Q_0$ together with $l_n,\;n>1$ satisfy the axioms of an $L_\infty$-algebra. 

We end this section by noting that all that was said in this section applies as well to world-sheets with boundaries. In particular, if we consider a disk with all punctures on its boundary, the above construction leads to a string field theory of open strings. What changes, is that the dimension of the relevant moduli spaces is half of that for closed strings, in particular for the disk with $n$ punctures the real dimension of ${P}_{n}$ is $n-3$. Also, while in closed string field theory we have invariance under permutations, in the open string one has invariance under rotations of the punctures on the boundary. Consequently, instead of an $L_\infty$-algebra, the vertices of open string field theory satisfy the axioms of an $A_\infty$-algebra.

\section{Open superstring field theory}\label{sec:4}

Superstring field theory is based on the moduli space\footnote{For simplicity we will consider only NS punctures in this talk.} $\hat{\mathcal{P}}_n$ for super Riemann surfaces of (even | odd) dimension $(3g-3+n|2g-2+n)$.  Unfortunately, we do not know what the generalization of the geometric master equation (\ref{BVgcq}) to $\hat{\mathcal{P}}_n$ is. So, we are faced with the problem of constructing a perturbative string field theory action on $V$ where neither the BV action (Section~\ref{sec:2}) nor the geometric BV equation on the moduli space (Section~\ref{sec:3}) is available. What comes to our rescue is the fact, described above, that, whatever the geometric BV equation is, it will translate into an $A_\infty$-structure on $V$. Furthermore, it turns out that, as a result of the equivalence of integration over an odd variable, $\eta$ and differentiation with respect to $\eta$, the chain map property allows to construct all higher order vertices (or maps) algebraically. 
 
The chain map is realized in analogy with the bosonic string as  
\begin{align}
\alpha_{x_1,\ldots,x_{m+3}}(h,\delta h)=\langle e^{-i<\delta h, b>-i<\delta\chi,\beta>}\prod\limits_{j=1}^{m+3}( f^*_j\mathbf{V}_j)(0)\rangle
\end{align}
where, in addition to the the conformal structure, $h$ we introduced the world-sheet gravitino $\chi$ and its variation $\delta\chi$  of even parity, and which is contracted with $\beta$, the superpartner of $b$. In what follows, we will consider open strings. The cubic vertex is given by a disc with 3 punctures. In that case the dimension of $\hat{\mathcal{P}}_3$ is $(0|1)$, that is, there is a single odd modulus and no even modulus. The cubic superstring vertex, $M_2$, is then obtained by integrating $\alpha$ over $\Pi T{\cal{\nu}}_3$, that is,
\begin{eqnarray}\label{m2tm}
\int_{\Pi T{\cal{\nu}_3}}\alpha_{\{x_i\}}(\eta, \dd \eta)&=&\omega(x_1,M_2(x_2,x_3))
\end{eqnarray}

Now, while $\eta$ and $\dd\eta$ are are good coordinates on $\Pi T{\cal{M}}$, they are somewhat misaligned with the action of the BRST charge, $Q_0$ in the CFT. In particular, while $\alpha_{\{x_i\}}(\eta, \dd \eta)$ defines a chain map from End$_V$ to $\Pi T{\mathcal{P}}_3$, the integral over the fibre
\begin{equation}
\tilde\alpha_{\{x_i\}}(\eta)=\int \dd(\dd \eta) \;\alpha_{\{x_i\}}(\eta,\dd\eta)
\end{equation}
is not a chain map from End$_V$ to ${\mathcal{P}}_3$ in contrast to the bosonic string in Section~\ref{sec:3}.  However, this can be cured by integrating over a singular fibre keeping 
\begin{eqnarray}
\tilde\eta =\frac{\eta}{\dd\eta}\,
\end{eqnarray}
fixed. With this choice $\tilde\alpha_{\{x_i\}}(\tilde\eta)$ has the desired chain map property since 
\begin{equation}
\begin{aligned}
 \int\frac{1}{{\dd\eta}} \alpha_{(Q_0 x_i)}(\tilde\eta,\dd\eta)\;\dd({\dd\eta})  &=  \int\frac{1}{{\dd\eta}} \dd\alpha_{(x_i)}(\tilde\eta,\dd\eta)\dd({\dd\eta})\\
 &=\dd\int\frac{1}{{\dd\eta}}\alpha_{(x_i)}(\tilde\eta,\dd\eta)\dd({\dd\eta})\,,
 \end{aligned}
\end{equation}
with $ \dd =\frac{\partial}{\partial\tilde \eta}$. Furthermore, since $\int \dd\tilde\eta\;f(\tilde\eta)=\dd\;f(\tilde\eta)|_{\tilde\eta=0}$ we find that $M_2$ is exact, 
\begin{equation}
M_2=[Q_0,\mu_2]
\end{equation}
where
\begin{equation}
\omega(x_1,\mu_2(x_2,x_3))=\tilde\alpha_{\{x_i\}}(0)\,.
\end{equation}
This, in turn, allows us to solve the $A_\infty$-relation 
\begin{eqnarray}
[Q_0,M_3]+\frac12[M_2,M_2]=0\,,
\end{eqnarray}
by $M_3=[\mu_2,M_2]$ up to an exact contribution. This procedure can then be applied recursively to determine all higher maps recursively. Furthermore, it can be shown that this construction is unique (see \cite{Erler:2013xta,Erler:2014eba,Erler:2015lya,Konopka:2016grr,Chiaffrino:2018jfy} for more details). To summarize, what we found is that the algebraic structure implied by the underlying BV structure completely determines the action of superstring field theory, even though the precise formulation of the geometric BV structure on super moduli space in unexplored so far. A note of caution is in order: while a BV structure on $\mathcal{F}$ always implies an $\infty$ structure on $T_p\mathcal{F}$, the converse is not guaranteed. Thus not every $\infty$ structure necessarily descends from a BV structure. 

\bibliography{allbibtex}
\bibliographystyle{prop2015}

\end{document}